\documentclass{article}
\usepackage{amsmath}
\usepackage{amssymb}
\usepackage{authblk}

\usepackage{graphicx}
\usepackage{dcolumn}
\usepackage{bm}
\usepackage{amsfonts}
\usepackage{color}

\usepackage[normalem]{ulem}

\begin{document}

\title{Share, but unequally: A plausible mechanism for emergence and maintenance of  intratumor heterogeneity}
\author[1]{Xin Li\thanks{xinlee0@gmail.com}}
\author[1]{D. Thirumalai\thanks{dave.thirumalai@gmail.com}}
 \affil[1]{Department of Chemistry, University of Texas at Austin, Texas 78712, USA}

\date{\today}

\maketitle

\noindent \textbf{Abstract}
\vspace{3mm}

\noindent Intratumor heterogeneity (ITH), referring to coexistence of different cell subpopulations in a single tumor, has been a major puzzle in cancer research for almost half a century. The lack of understanding of the underlying mechanism of ITH hinders  progress in developing effective therapies for cancers.  Based on the findings in a recent quantitative experiment on pancreatic cancer,  we developed a general evolutionary model  for one type of cancer, accounting for interactions between different cell populations through paracrine or juxtacrine factors.  We show that the emergence of a stable heterogeneous state in a tumor requires an unequal allocation of paracrine growth factors (``public goods'') between cells that produce them and those that merely consume them.   Our model provides a quantitative explanation of recent {\it in vitro} experimental studies in pancreatic cancer in which insulin growth factor (IGF-II) plays the role of public goods.  The calculated phase diagrams as a function of exogenous resources  and fraction of growth factor producing cells show ITH persists only in a narrow range of concentration  of exogenous IGF-II. Remarkably, maintenance of ITH requires cooperation among tumor cell subpopulations in harsh conditions, specified by lack of exogenous IGF-II, whereas surplus exogenous IGF-II elicits competition. Our theory also quantitatively accounts for measured {\it in vivo} tumor growth in glioblastoma multiforme (GBM).  The predictions for GBM tumor growth as a function of the fraction of tumor cells are amenable to experimental tests. The mechanism for ITH also provides hints for devising efficacious therapies. \\




\section*{Introduction}  

Cancer, a complex disease that arises through clonal evolution, is a major cause of mortality throughout the world with no cure in sight despite a tremendous amount of effort and resources expended to uncover its root causes. The underlying  mechanisms of the origin and spread of  cancer is still under debate~\cite{Gerlinger2014}. The first evolutionary theory of cancer, proposed by Nowell in 1976,  describes cancer progression as a linear process derived from  sequential acquisition of somatic mutations~\cite{Nowell1976}.  With the advent of next generation sequencing~\cite{Vogelstein13} it has become clear that instead of linear growth cancer evolution is better described by branched growth in  which multiple subclones appear and coexist during cancer progression. Accumulating evidence favor the branched model and the associated intratumor heterogeneity (ITH) ~\cite{Gerlinger2012,Sottoriva2013,Bashashati2013,Gerlinger20142,Bruin2014,Yates2015,Boutros2015,Ling2015}.  ITH is a complex phenomenon and  many sources, such as genetic, epigenetic mutations, stochastic genetic expression and so on, could contribute to its origin~\cite{Almendro13}. The presence of ITH  in a variety of cancers, which is a great impediment in designing effective treatments~\cite{Almendro13,Bedard2013,Zhang2014}, is hard to rationalize  according to the linear evolutionary model because subclones with even a small fitness advantage should  ultimately proliferate at the expenses of others.  For this reason the persistence of ITH  in a macroscopic tumor is a puzzle.

The tumor cells, with diverse genetic or epigenetic mutations, are often spatially separated~\cite{Gerlinger2012,Gerlinger20142} with each subclone dominating the cell population in a specific region. It indicates that spatial constraints or distinct microenvironments might prohibit clonal sweeps, thus inducing  ITH.  However, it cannot explain the coexistence of distinct subclones in the same region of a tumor~\cite{Snuderl2011,Szerlip2012}.

The interactions between tumor cells and the surrounding normal cells,  and microenvironments have been extensively studied in the past few decades~\cite{Liotta2001,Allinen2004,Bhowmick2004,malmi2017cell} while much less effort has been made in investigating the interactions among subclonal populations in tumors. 
Instead of competition, the cooperation among  distinct subclonal populations is found to be essential for tumor maintenance~\cite{Cleary14}, enhance tumor growth~\cite{Marusyk2014}, and even facilitate cancer metastasis~\cite{Chapman2014,Aceto2014}. Meanwhile, it is  observed that  a minor and even undetectable subclone can  dominate the clinical course and lead to cancer relapse frequently~\cite{Mullighan2008,Johnson2014,Morrissy2016}. Therefore, it is crucial to understand the mechanism of cooperation that facilitates the emergence and maintenance of ITH in a single tumor.  

Cooperation can be established through mutualism or even unidirectional interactions among different subpopulations. Mutualism in ecology provides an effective mechanism for the establishment of a heterogeneous state in which each subpopulation benefits from the activity of the other~\cite{Boucher1982,menon2015public,zhou2018circuit,wang2011asymmetric}.  Recently, it was found that two distinct subclones in cancer can complement each other's deficiency in order to survive and proliferate~\cite{Cleary14}. The formation of a heterogeneous state can be explained by a mechanism similar to mutualism in which fitness of distinct cell types is maximized by resource sharing~\cite{Axelrod2006}.  Compared to the strict interdependence in mutualism, a unidirectional interaction between distinct populations is observed more frequently~\cite{Axelrod2006,merlo2006cancer,Inda2010}. It is quite common to find that some tumor cells secrete diffusible growth factors or other paracrine factors to promote tumor growth~\cite{Axelrod2006}. Meanwhile, other types of tumor cells free ride on those essential nutrients to grow and might even dominate the whole population without producing them.  One recent study for Glioblastoma multiforme (GBM)  shows that a minor subclone in a tumor supports and promotes the growth of a dominant one by activating a paracrine signaling circuit~\cite{Inda2010}. It was  found that the mixture of these two distinct subpopulations promote faster tumor growth  than they would by themselves.  Similarly, an insightful {\it in vitro} experiment combined with theoretical analysis based on evolutionary game theory~\cite{Archetti15} investigated the ``public goods game" in pancreatic cancer cell populations in which one cell type produces a growth factor as the ``public good". The growth factors promote the proliferation of both cell types. It is found that  these two types of cancer cells can coexist under certain conditions although there is only a unidirectional interaction between them.  Although insightful, previous applications of  the evolutionary game theory provided only a qualitative explanation for the experimental results~\cite{Archetti15}. The assumption that the population size is a constant does not accurately capture the growth dynamics observed in their experiments. In addition, as shown here, such an assumption cannot account for the intriguing related phenomena of glioblastoma multiform growth~\cite{Inda2010} in which there are clear manifestations of ITH. 
Additionally, a theoretical framework accounting for influences of different factors such as exogenous resources and the cost for producing public goods on the establishment of cooperation between two cell populations directly is needed for expanding the scope of the game theory applications. Therefore, the underlying generality of the mechanisms for the origin of ITH is still unclear.


Here, we investigate the mechanism of cooperation between two distinct populations composed of `producers' and `non-producers'.  We show that several factors are indispensable for the maintenance of a stable heterogeneous coexisting population of producer and non-producer cells.  A critical finding in our work is that the unequal allocation of  public goods among different species plays a crucial role in maintaining  heterogeneity.  By studying the influence of exogenous resources and initial population fraction on such a simple system, we obtained a phase diagram from our theory, which is in excellent agreement with experimental observations. Our theory also quantitatively explains the unexpected growth behaviors of  GBM tumors as a function of initial fractions of producer cells in  {\it in vivo} experiments with no free parameters.  The robustness of the theory is established by making testable predictions of the origin of ITH in GBM  driven by a paracrine mechanism~\cite{Inda2010}.  The discovery of mechanisms for  such ITH might also give clues  for changing strategy  of  treatment in cancers in which interactions between  different subclones are prevalent.


\section*{Models}

\begin{figure}
\vspace*{-0.1cm}
\hspace*{-0.2cm}
\centering
\includegraphics[clip,width=1.0\textwidth]{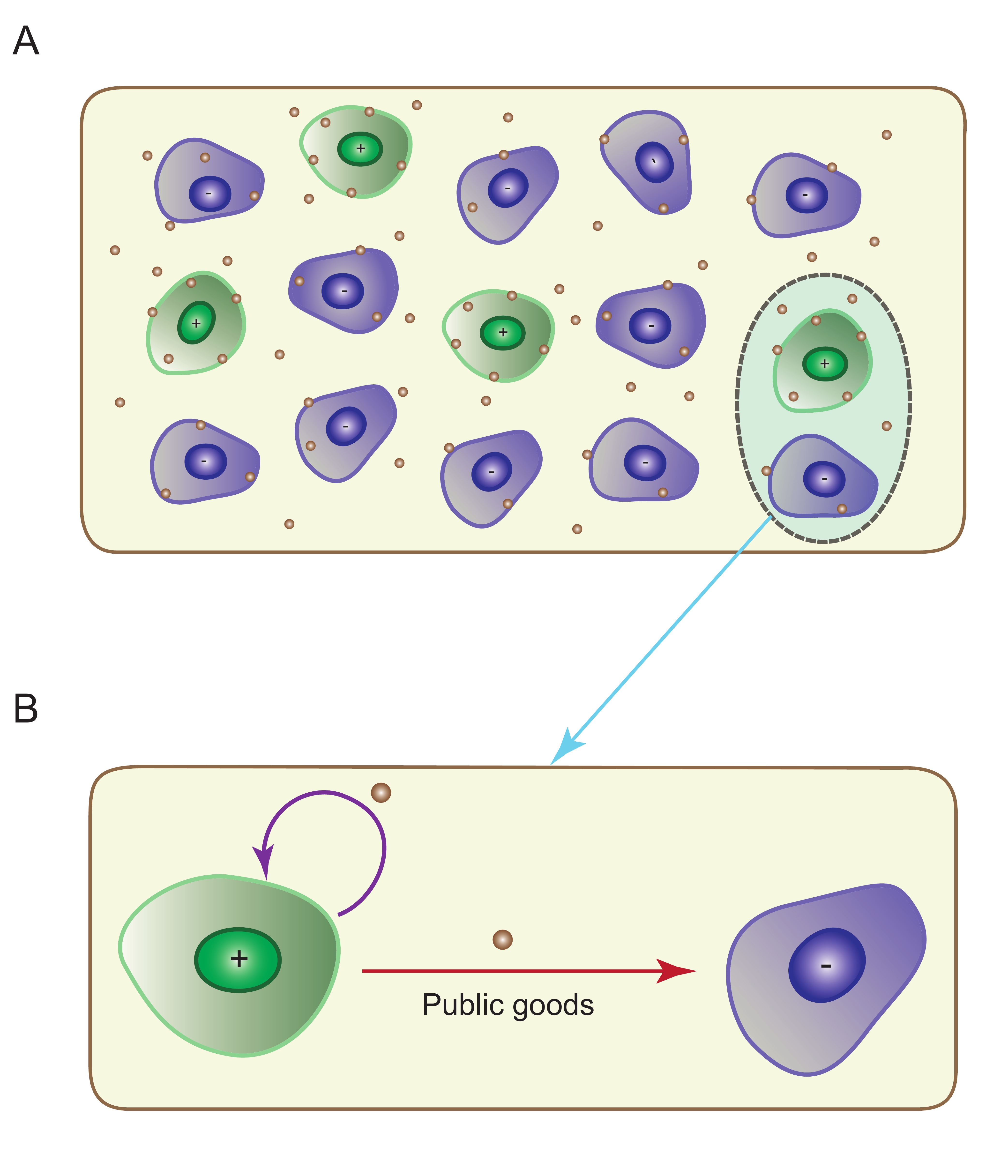}
\caption{\label{fig:schematicfig} {\bf Illustration of public goods game.} (A) The public goods (small brown circles) generated by producers (``+'' agents in green) are shared  unequally between producers and non-producers (``-'' agents in blue color). Both producers and non-producers benefit from the presence of the public goods, which promote proliferation or survival of these agents. The public goods can also be supplied from exogenous resources. (B) A zoom-in  of the dashed line oval in Fig.~\ref{fig:schematicfig}A to illustrate the public goods dependent circuit for the producer and non-producer. Coexistence of the two cell types requires feedback (purple line) and unequal sharing of the public goods.}
\end{figure}

The public goods game has been extensively used in the studies of human societies, and similar concepts have been applied to other systems, such as microbial colonies, and insect communities~\cite{Hofbauer1998,Hauert2006,nowak2006five,Allen2013,Nanda2017,bauer2018role}. In this model, both  the  producer and non-producer benefit from the products produced solely by the former (see Fig.~\ref{fig:schematicfig}). However, the producers pay a price for the production of public goods while the non-producers merely free-ride on the products without incurring any cost. In general, such a dynamics would result in an unstable system in the sense the producer might become extinct if the public goods are shared equally between the parties. The system would collapse unless sufficient exogenous resources are provided, which could lead to the fixation of non-producers, as discussed below in detail. 

{\bf Equal sharing is untenable:} 
We describe the evolution of the fraction $f_{+}$ of producers and $f_{-}$ of non-producers using the replicator equations,
\begin{equation}
\label{f+}
\frac{\partial f_{+}}{\partial t} = (w_{+}-\langle w \rangle)f_{+}\, , 
\end{equation} 
and,
\begin{equation}
\label{f-}
\frac{\partial f_{-}}{\partial t} = (w_{-}-\langle w \rangle)f_{-}\, ,  
\end{equation} 
where $w_{+}$ and $w_{-}$ are fitness of producers, and non-producers, respectively. The normalization condition is $f_{+} + f_{-} = 1$. The average fitness $\langle w \rangle$ is,
\begin{equation}
\label{fitaverage}
\langle w \rangle = w_{+} f_{+} + w_{-}f_{-}\, . 
\end{equation} 
Let $N_{+}$, and $N_{-}$ be the number of producers and non-producers, respectively. The total population size is $N\equiv N_{+}+N_{-}$. Although the system size, $N$, is  often assumed to be a constant~\cite{Archetti15,Blythe07}, we consider a general case~\cite{Melbinger2010} where the population size varies with the time evolution given by,
\begin{equation}
\label{Number}
 \frac{\partial N}{\partial t} =w_{+}N_{+}+w_{-}N_{-} = \langle w \rangle N \, . 
\end{equation} 
Agent death, neglected for simplicity, could be readily included in the fitness functions.

If the public goods are shared among all the producers and non-producers equally, the same benefit will be presented to them,  leading to the relation,
\begin{equation}
\label{wrelation}
w_{+} = w_{-} - p_{0} \, , 
\end{equation} 
where  $p_{0}$ ($> 0$) is the cost paid by the producers to generate the public goods. Then, the time evolution of producers could be rewritten as,
 \begin{equation}
\label{f+r}
\frac{\partial f_{+}}{\partial t} = -p_{0}f_{+}(1-f_{+}) \, . 
\end{equation} 
Therefore, the fraction of producers would decrease with time until it vanishes because  $p_{0}$, $f_{+}$, and $1 - f_{+}$ are  all  non-negative. Then, the non-producers could  sweep through the population, achieving higher fitness in the process, if exogenous public goods are provided continuously to support the population growth. Otherwise, the system would collapse as observed in the case of ``tragedy of the commons" once the public goods are depleted~\cite{Hardin1968}.

\section*{Results}

{\bf Coexistence between producers and non-producers requires non-linear fitness:} A solution to the dilemma that  emerges from the naive model of equal sharing, discussed above, is to change the rule for the allocation of public goods. Because the producer pays a price for the production of public goods, which decreases its fitness directly, more products should be allocated to the producer (see the public goods distribution in Fig.~\ref{fig:schematicfig}A as an example). If this were to occur the producer would recoup the losses in order to gain the same fitness as the non-producer. In this unequal sharing scenario, the producer and non-producer could coexist, as we show here. 

In general, the fitness of one agent is a function of the fraction of producers. A higher fitness is expected as the fraction of producers increases. For a dynamic system, a heterogeneous state (containing both producers and non-producers) could be maintained only  the stabilities of the states are ensured. A heterogeneous state cannot be realized if $w_{+}$ and $w_{-}$ are  linear fitness functions of $f_{+}$. Let the fitness functions of the producer ($w_{+}$) and non-producer ($w_{-}$) be,
\begin{equation}
\label{linear+}
w_{+} = k_{+}f_{+}- p_{0} \, , 
\end{equation} 
and 
\begin{equation}
\label{linear-}
w_{-} = k_{-}f_{+} \, , 
\end{equation} 
respectively. The parameters $k_{+}$, $k_{-}$ are the corresponding allocation coefficients of public goods produced by producers and $p_{0}$ is the cost paid by each producer. The producer needs to get more public goods than the non-producer, which means $k_{+} > k_{-}$ ($k_{+} = k_{-}$ in Eq.~(5)). The condition for  equilibrium  with both players follows from Eqs.~(1) and (2),  resulting in,
\begin{equation}
\label{linearsol}
 f_{+}^{0}=\frac{p_{0}}{k_{+}-k_{-}} \, . 
\end{equation} 
Given the fraction $f_{+}^{0}$ for the producer in Eq.~(\ref{linearsol}), the two members attain the same fitness ($w_{+}=w_{-}$). However, this equilibrium is unstable and eventually only one of them survives (see Fig.~S1 in the SI). As the fraction $f_{+}$ becomes a little higher than $f_{+}^{0}$ due to the birth of new producers or a higher value is given initially, the producer would attain higher fitness than the non-producer (for $k_{+}>k_{-}$), leading to a much higher fraction of  the producer. Finally, the producer would take over the whole population due to this positive feedback. In the opposite limit, the system would consist of non-producer only if $f_{+}$  is smaller than $f_{+}^{0}$. Therefore, a linear fitness function cannot lead to the establishment of a stable heterogeneous system in the present model. Instead of a linear function, fitness functions are typically non-linear in biological systems at all length scales due to cooperation or competition between the various interacting moieties~\cite{chesson2000mechanisms,hauert2006synergy,archetti2013dynamics,kimmel2018neighborhood}. In the following, we first utilize data from one recent experiment to illustrate how a stable heterogeneous population can be established from the  public goods game in a system consisting of both producers and non-producers where a non-linear relation is observed for fitness functions~\cite{Archetti15}.

In a recent study,  Archetti, Ferraro and Christofori (AFC) investigated the origin of the``tragedy of the commons" in cancer cells~\cite{Archetti15}. The insulin-like growth factor II (IGF-II) is up-regulated in many cancers, which can promote cell proliferation and abrogate apoptosis~\cite{Christofori94,Pollak08}. The producer (+/+) cells are derived from mice with insulinomas (a neuroendocrine pancreatic cancer), while the non-producer (-/-) cells are obtained from the same mice but with IGF-II gene deleted. Therefore, the -/- cells do not produce the IGF-II molecules. Thus, IGF-II is an ideal public good for these two cell types because both of them can uptake this protein to reach higher fitness (growth rate), ensuring their survival and growth. The two different types of cells were then mixed  to investigate the conditions under which a stable heterogeneous state could be established, mediated by optimal sharing of IGF-II.  Although AFC proposed a sound analyses of their findings based on game theory, only qualitative comparisons to their experiments were provided. In addition, the assumption of a constant population size~\cite{gerlee2015complexity} also requires scrutiny.  Here we approach the problem from a different perspective relying on the replicator dynamics with evolving population size, which enables us to make quantitative predictions  not only for pancreatic cancer but also GBM.

The measured growth rate of the non-producer -/- cells as a function of exogenous IGF-II concentration, $c$, is non-linear (see Fig.~S2 in the SI). In order to solve the replicator equations, we first fit the  experimentally measured -/- cell growth rate using a Hill-like function,
\begin{equation}
\label{fit-ex}
w_{-} = a_{1} + \lambda_{1}c^{\alpha}/(a_{2}^{\alpha}+c^{\alpha}) \, .
\end{equation} 
The Hill function in Eq.~(\ref{fit-ex}) fits the experimental data accurately (see Fig.~S2 in the SI) yielding $a_{1} = 2.0$, $\lambda_{1} = 18.9$, $\alpha = 0.7$, and $a_{2} = 3.2$. 
We also used the logistic function, which does not fit the data nearly as well,  to make predictions (see Figs.~S3, S4 and SI for details).
Interestingly, the use of logistic function for the fitness yields qualitatively similar results (see Figs.~S3, S4 and the SI for details), thus demonstrating that non-linear feedback between +/+ and -/- cells is the source of heterogeneity, as already surmised by AFC.  We describe the results in the rest of the paper using  the Hill-like function for $w_{+}$ and $w_{-}$. 

The public good IGF-II is produced endogenously or can be supplied exogenously. Therefore, we write  the IGF-II ($c_{-}$) available for the non-producer as, 
\begin{equation}
\label{IGFIIC}
 c_{-}\equiv c = b f_{+} + c_{0} \, . 
\end{equation} 
where $c_{0}$ represents the exogenous supply of IGF-II,  and $b f_{+}$ is the allocation of IGF-II arising from +/+ cells. Because AFC did not measure the fitness of the producer cells systematically, a  relation similar to that in Eq.~(\ref{fit-ex}) for $w_{+}$, might be assumed.   In order for the emergence of heterogeneous populations, the allocation of public goods produced by the +/+ cells should be unequal, so the growth rate of +/+ cells is written as,
\begin{equation}
\label{fit+ex} 
w_{+} = g(c_{+}) -p_{0} \, ,
\end{equation} 
where $g(c_{+})$ has the same Hill-like functional form as in Eq.~(\ref{fit-ex}) except that $c$ is replaced by $c_{+}$,  leading to, 
\begin{equation}
\label{cplus} 
c_{+} = af_{+}+c_{0} \, .
\end{equation} 
The parameter $a$, similar to $b$ in Eq.~(\ref{IGFIIC}), is the coefficient for the allocation of IGF-II produced by +/+ cells.

\begin{figure}
\vspace*{-0.9cm}
\hspace*{-0.2cm}
\centering
\includegraphics[clip,width=0.95\textwidth]{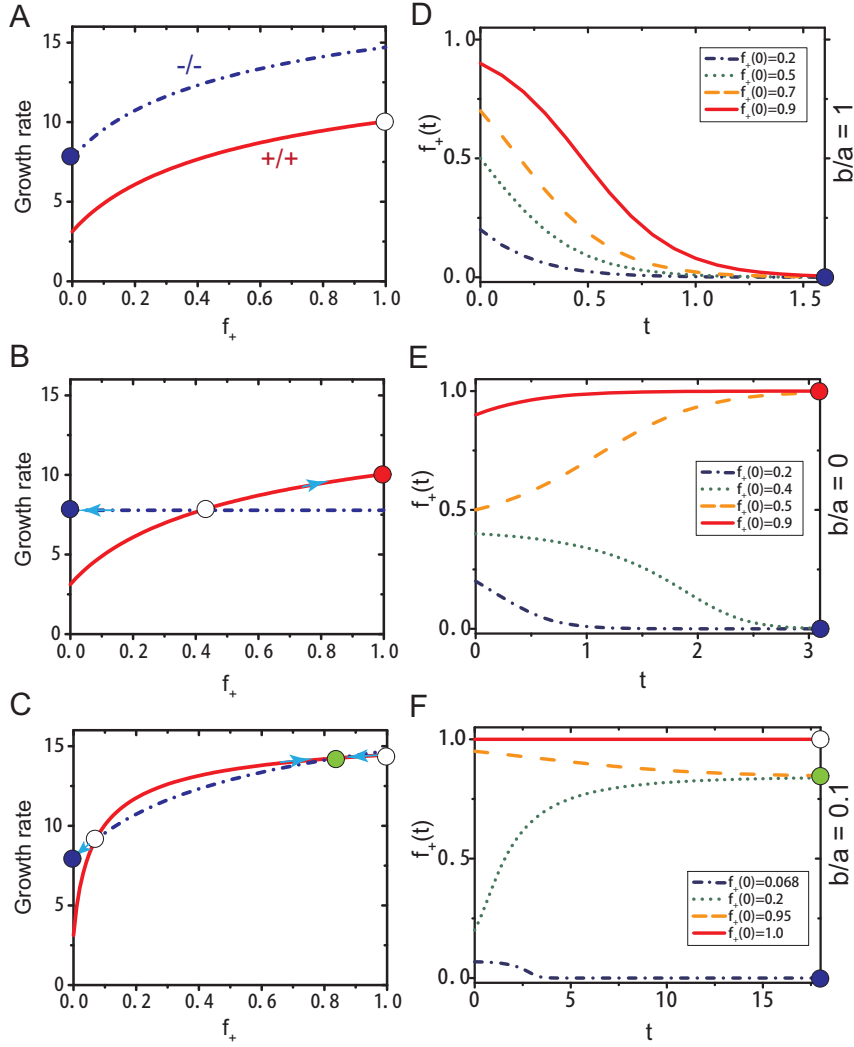}
\caption{\label{fig:Share} (A)-(C): Growth rates of +/+ and -/- cells as a function of the fraction ($f_{+}$) of +/+ cells under different allocation of IGF-II produced by the +/+ cells. (A) Equal share of IGF-II ($b=a=8$), (B) no share ($b=0$, and $a=8$), (C) a small portion ($b=8$, and $a=80$) is allocated  to -/- cells. The value of $c_{0} =1$ and $p_{0}= 4.65$ in Eq.~(\ref{fit+ex}) are derived from fitting the equilibrium fractions of +/+ cells observed in experiments using our model. The growth rate of +/+ cells are shown in solid red lines while dash-dotted blue lines describe the growth rate of -/- cells. The filled and empty circles  indicate a stable or unstable fixed point, respectively. A  stable state consisting of only +/+ (-/-) cells is indicated in red (blue) color. The  green filled circle shows a stable heterogeneous state representing coexistence of the two cell types. (D)-(F):  The evolution of $f_{+}(t)$ at different $f_{+}(t=0)$ values corresponding to the allocation strategies of IGF-II in (A)-(C). Each row represents results from one of the three allocation strategies.  The growth rate is defined as the relative density change of cells during the log phase~\cite{Archetti15}. The unit for time is days.}
\end{figure}

\textbf{Influence of public goods allocation strategies:} First, let us consider the influence of  allocation strategies for public goods in a mixture of +/+ and -/- cells. The ratio of $b/a$ in Eqs.~(\ref{IGFIIC}) and (\ref{cplus}) determine how  the public goods provided by the producers are shared between the two populations. The public goods are shared equally if the ratio ($b/a$) is equal to unity, while the producers do not provide resources to the non-producer if $b/a = 0$.  More resources are allocated to the producer if $b/a <1$ while  the non-producer obtains a larger amount of resources as $b/a >1$.    Public goods (IGF-II) is diffusible, which is  modeled in our theory in the following way. The equality ($a = b$) would result if diffusion of IGF-II is rapid. On the other hand, the inequality, $a > b$, would be a consequence of slow diffusion or fast uptake of IGF-II by the producers. The assumption of slow diffusion, with a limited diffusion range of IGF-II, is made in the model of AFC~\cite{Archetti15}.  Thus, by considering different values of the ratio, $b/a$, different rates of diffusion and uptake of IGF-II are covered. Accordingly, three values 1, 0,  and 0.1 are considered for the ratio $b/a$ in Figs.~\ref{fig:Share}A-C, which show the growth rate of the two cell types as a function of the fraction ($f_{+}$) of +/+ cells. The corresponding evolution of $f_{+}(t)$ under different initial conditions are also shown in Figs.~\ref{fig:Share}D-F. 

The -/- cells  always grow faster than +/+ cells if the public goods are shared equally ($b/a=1$) (see Fig.~\ref{fig:Share}A), which  can also be derived from Eqs.~(\ref{fit-ex}) and (\ref{fit+ex}) directly ($w_{-}>w_{+}$ for $b\geq a$). The non-producer always attains higher fitness than the producer as long as  the former gets larger ($b/a > 1$) or equal ($b/a = 1$) amount of IGF-II.  Therefore,  the non-producer would take over the whole population if $b/a \geq 1$ producing a homogeneous state, irrespective of the initial fraction $f_{+}(0)$ of the producer  (see the evolution of $f_{+}(t)$ in Fig.~\ref{fig:Share}D). The exception is when $f_{+}(0) = 1$. The state with producers only ($f_{+} = 1$, see the open circle in Fig.~\ref{fig:Share}A) is unstable under infinitesimal perturbations of non-producer population. A steady binary system with the coexisting population of +/+ and -/- cells cannot  form under these conditions, as expected from previous arguments. 

Consider another limiting case with $b/a = 0$ in which  the non-producer does not get access to the public goods generated by the producer (with $w_{-} =$ constant, see the dash-dotted blue line in Fig.~\ref{fig:Share}B). In this limit, it is possible to have an internal equilibrium (see the open circle in Fig.~\ref{fig:Share}B) resulting in  the two cells having the same fitness ($w_{-} = w_{+}$). However, this is an unstable equilibrium state, which means only one type of population would survive (see the filled red and blue circles in Fig.~\ref{fig:Share}B, respectively) depending on the initial conditions (see also the evolution of $f_{+}(t)$ in Fig.~\ref{fig:Share}E).   A stable homogeneous state consisting of only producers results if $f_{+}(0)$ is above the fraction ($f_{+}^{us}$) of the producer corresponding to the internal unstable state (illustrated by the open circle in Fig.~\ref{fig:Share}B).  In this case, the flow is towards the $f_{+}(0)=1$ stable state. For $f<f_{+}^{us}$, the non-producers form another  stable homogeneous state (see the green dotted and blue dash-dotted lines in Fig.~\ref{fig:Share}E). A stable heterogeneous state with coexisting populations cannot exist if $b/a=0$.

If the public goods are allocated unequally between the +/+ and -/- cells  due to the presence of spatial structure, (Fig.~\ref{fig:Share}C) with $0< b/a <1$, two internal equilibrium states  (with $0< f_{+}^{i}< 1$) appear. One of them is an unstable state (the left open circle in Fig.~\ref{fig:Share}C) whereas the  filled green circle in Fig.~\ref{fig:Share}C corresponds to a stable state due to the frequency-dependent selection~\cite{Archetti15,ayala1974frequency}. Close to the internal stable state,  the fitness of +/+ cells becomes smaller than that of -/- cells as the  +/+ cell frequency increases above the stable state value. Therefore, the frequency of +/+ (-/-) cells will decrease (increase) until it returns to the value corresponding to the stable state (see the yellow dashed lines in Fig.~\ref{fig:Share}F). Similar effect is observed when the +/+ cell frequency is below the stable state value (see the green dotted lines in Fig.~\ref{fig:Share}F) as long as the $f_{+}(0)$ is higher than the value $f_{+}^{us} = 0.069$ corresponding to the internal unstable state.  Therefore, the two types of cells coexist, leading to  a stable heterogeneous state. We also observe that  producers would be extinct (see the blue dash-dotted line in Fig.~\ref{fig:Share}F) if a small amount of +/+ cells are mixed  with a large population of -/- cells initially. These findings based on the replicator equations with non-linear $w_{-}$ are consistent with the  experimental observations obtained in the presence of a small amount of exogenous resources~\cite{Archetti15}.

\begin{figure}
\vspace*{-0.1cm}
\hspace*{-0.2cm}
\centering
\includegraphics[clip,width=1.0\textwidth]{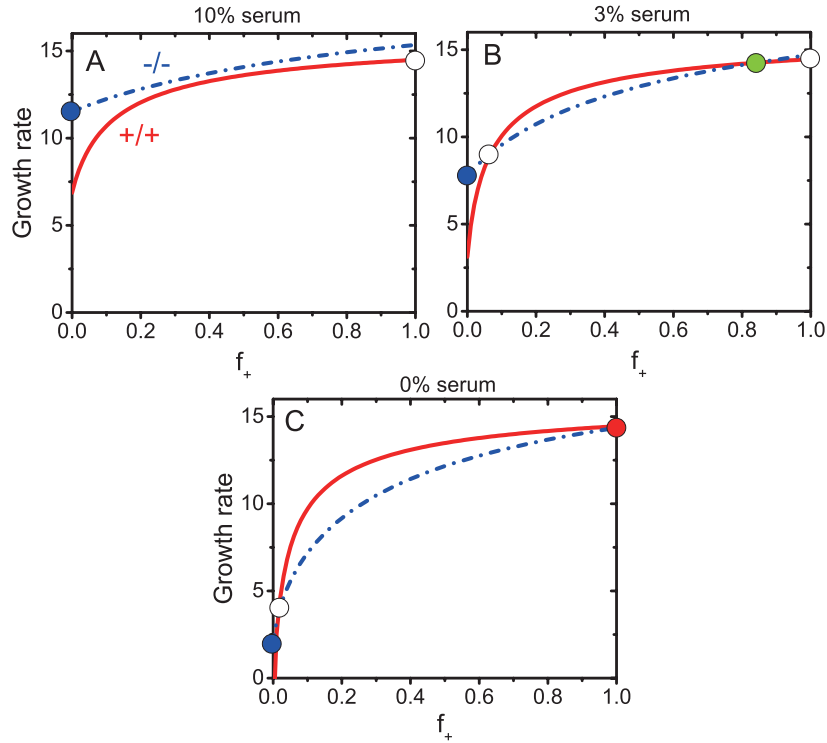}
\caption{\label{fig:Serum} Growth rates of +/+ and -/-  cells as a function of the fraction ($f_{+}$) of +/+ cells  at different  levels of exogenous resources (serum).  (A) $c_{0} = 3.3$ corresponds to $10 \%$ serum in experiments; (B) $c_{0} = 1$ represents $3 \%$ serum; (C) $c_{0} = 0$ implies absence of serum. The value of $a = 80, b = 8$, and $p_{0}= 4.65$, corresponding to the parameters in Fig.~\ref{fig:Share}C. The flow diagram in Fig.~\ref{fig:Serum}B corresponds to Fig.~\ref{fig:Share}C.  The meaning of the symbols used and the definition of growth rate are the same as in Fig.~\ref{fig:Share}.}
\end{figure}

\textbf{Role of the exogenous production of public goods:} From Eq.~(\ref{IGFIIC}), it is clear that the stable state could also be influenced by extrinsic factors, such as the  supply of exogenous resources.    By varying the values of the parameter $c_{0}$ in Eqs.~(\ref{IGFIIC}) and (\ref{cplus}), we investigated the role of exogenous resources in public goods game. The growth rate of the two different types of cells  as a function of the fraction of +/+ cells is shown in Fig.~\ref{fig:Serum} as the supply of exogenous resources (serum in the experiments~\cite{Archetti15}  containing IGF-II molecules) is changed. 

Fig.~\ref{fig:Serum}A shows that -/- cells  grow faster than +/+ cells irrespective of the fraction $f_{+}$ of +/+ cells given large enough exogenous public goods (large $c_{0}$). Surprisingly, we find that the non-producer would sweep through the population while the producer becomes extinct (see the blue filled circle in Fig.~\ref{fig:Serum}A) even though the latter could get additional public goods produced by themselves ($b/a \ll 1$). This is due to the competitive advantage of the non-producers in a high welfare environment without punishment.  These two cell types compete but do not cooperate with each other in nutrient-abundant environments.

As we decrease the serum concentration (smaller $c_{0}$) and keep all other parameters the same as in Fig.~\ref{fig:Serum}A, the two cell types start to establish a cooperative relationship and could coexist (see Fig.~\ref{fig:Serum}B). It leads to the appearance of a stable heterogeneous state (see the green filled circle in Fig.~\ref{fig:Serum}B), indicating that cooperation could be established more readily under harsh conditions (small available exogenous resources).   In addition, the whole system could attain higher fitness or drug resistance (see more detailed discussions later) due to the coexistence of both the  players.  It is well known that many cancer cells have to confront  hypoxia, low pH, low glucose and other severe conditions~\cite{Harris2002,Kato2013}, and they are often found to be more aggressive and dangerous compared to cancer cells under normal growth conditions with access to essential nutrients~\cite{Gupta2006}.  These conditions might make it favorable to  establish cooperation among them, leading to the formation of  heterogeneous populations. 

If exogenous resources removed from the system completely ($c_{0}=0$), the +/+ cells  dominate the whole population while the -/- cells would be swept away (see the red filled circle in Fig.~\ref{fig:Serum}C)  at sufficiently high $f_{+}(0)$. In the opposite limit ($f_{+}(0)$ is small), the -/- cells can take over the whole population (see the blue filled circle in Fig.~\ref{fig:Serum}C).  The phenomenon of tragedy of the commons will be observed if the public goods are essential for the survival of non-producers. Taken together these results show that the establishment of cooperation between different players shows strong dependence on environmental conditions. The influences of exogenous public goods as observed in Fig.~\ref{fig:Serum}  are consistent with the experimental results~\cite{Archetti15}, and are further discussed  below.

\begin{figure}
\vspace*{-0.1cm}
\hspace*{-0.2cm}
\centering
\includegraphics[clip,width=1.0\textwidth]{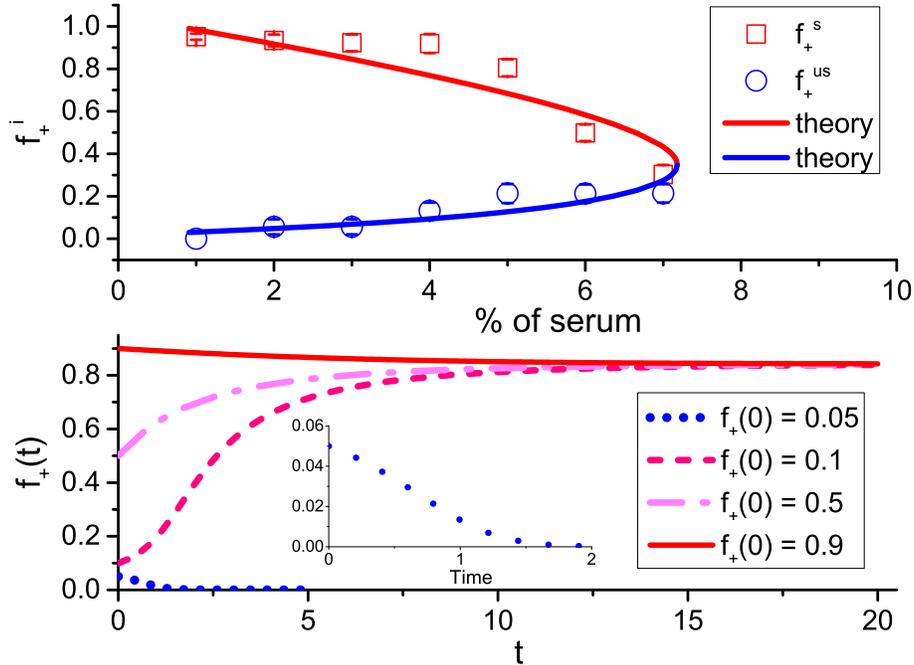}
\caption{\label{fig:phase} Upper panel: The internal equilibrium fractions ($f_{+}^{i}$, $i \equiv s$ or $us$, with $s$ for stable and $us$ for unstable state) of +/+ cells as a function of serum levels. Stable states are shown by red squares while blue circles indicate unstable states. The upper and lower bounds represent the upper and lower boundaries for the equilibrium fractions observed in experiments and the symbols give the middle value of these two boundaries. The solid lines correspond to theoretical predictions using $a = 80$, and $p_{0} = 4.65$ in Eqs.~(\ref{fit+ex}) and (\ref{cplus}). Lower panel:  Theoretical predictions for the time-dependent changes in the fraction $f_{+}(t)$ of +/+ cells for various initial conditions ($f_{+}(0)$ = 0.05, 0.1, 0.5, 0.9) in the presence of  $3\%$ of serum.  The inset figure shows the evolution of the fraction of +/+ cell with $f_{+}(0) = 0.05$.  The unit for time is days.}
\end{figure}

\textbf{Comparison with {\it in vitro} experiments:} Based on the calculations presented so far, we can now obtain the internal equilibrium fractions ($f_{+}^{i}$) of +/+ cells at different concentrations of serum. The experimental observations for $f_{+}^{i}$  are given by symbols  in the upper panel of Fig.~\ref{fig:phase}. The fractions $f_{+}^{s}$ ($f_{+}^{us}$)  under  stable (unstable) internal state are represented by red squares (blue circles). From experimental result (Fig.~1A) in \cite{Archetti15} for the equal fitness ($\approx 14.4$) of producer and non-producer cells and the fraction of +/+ cells approaching 1 for the  stable internal state at $c_{0}=0$, we obtain the parameter $b \approx 8$.  Our theoretical model with two free parameters $a$ and $p_{0}$ fits the experimental results very well (see the solid red and blue lines in the upper panel of Fig.~\ref{fig:phase}). One scale parameter has been used with $c_{0} = 1, 2,...$ corresponding to $3\%, 6\%,...$ of serum.

To illustrate the stability of the internal equilibrium state, an example is given in the lower panel of Fig.~\ref{fig:phase}, which describes the evolution of the fraction of +/+ cells under different initial conditions ($f_+(0)$). The serum amount is fixed at $3\%$.  It shows clearly that  a stable  equilibrium state is attained as long as $f_{+}(0)$ is above $f_{+}^{us}$, corresponding to the unstable internal equilibrium state. Then, the two types of cells cooperate leading to coexistence. However, the +/+ cells could be swept out and only -/-  cells exist if $f_{+}(0)$ is below $f_{+}^{us}$ (see the dotted line in the inset of the lower panel of Fig.~\ref{fig:phase}). From the results in the upper panel of Fig.~\ref{fig:phase}, it follows that there exists a critical concentration $c_{0}^c$ for exogenous resources (around 7$\%$ of serum in the experiment~\cite{Archetti15}).  A bistable system can be reached only if the concentration of serum is lower than $c_{0}^c$. One stable state corresponds to a heterogeneous system with two subpopulations and the other one is a homogeneous state consisting of only -/- cells (see Fig.~\ref{fig:Serum}B). In this scenario, maintenance of heterogeneous state is due to cooperation between +/+ and -/- cells. In contrast, as the concentration of serum increases beyond $c_{0}^c$, the -/- cells can always obtain sufficient resources to support a faster growth rate than +/+ cells (see Fig.~\ref{fig:Serum}A). Then, the -/- cell would sweep through the whole population as long as its initial fraction is non-zero.  Therefore, competition rather than cooperation is promoted between cell subpopulations under resource-abundant conditions.  It eventually leads to the establishment of a homogeneous system with only a single type of cell population. 

\textbf{Effects of price paid by producers:} In previous sections, we established that supply and allocation of public goods play crucial roles in determining the interactions among  subpopulations. We investigate the influence of another parameter $p_{0}$, the price paid by +/+ cells  to produce the public goods, which in the AFC experiment is IGF-II. The fitness of producers is affected by this parameter directly (see Eq.~(\ref{fit+ex})), so we anticipate that it will influence the relationship between producers and non-producers.

\begin{figure}
\vspace*{-0.1cm}
\hspace*{-0.2cm}
\centering
\includegraphics[clip,width=1.05\textwidth]{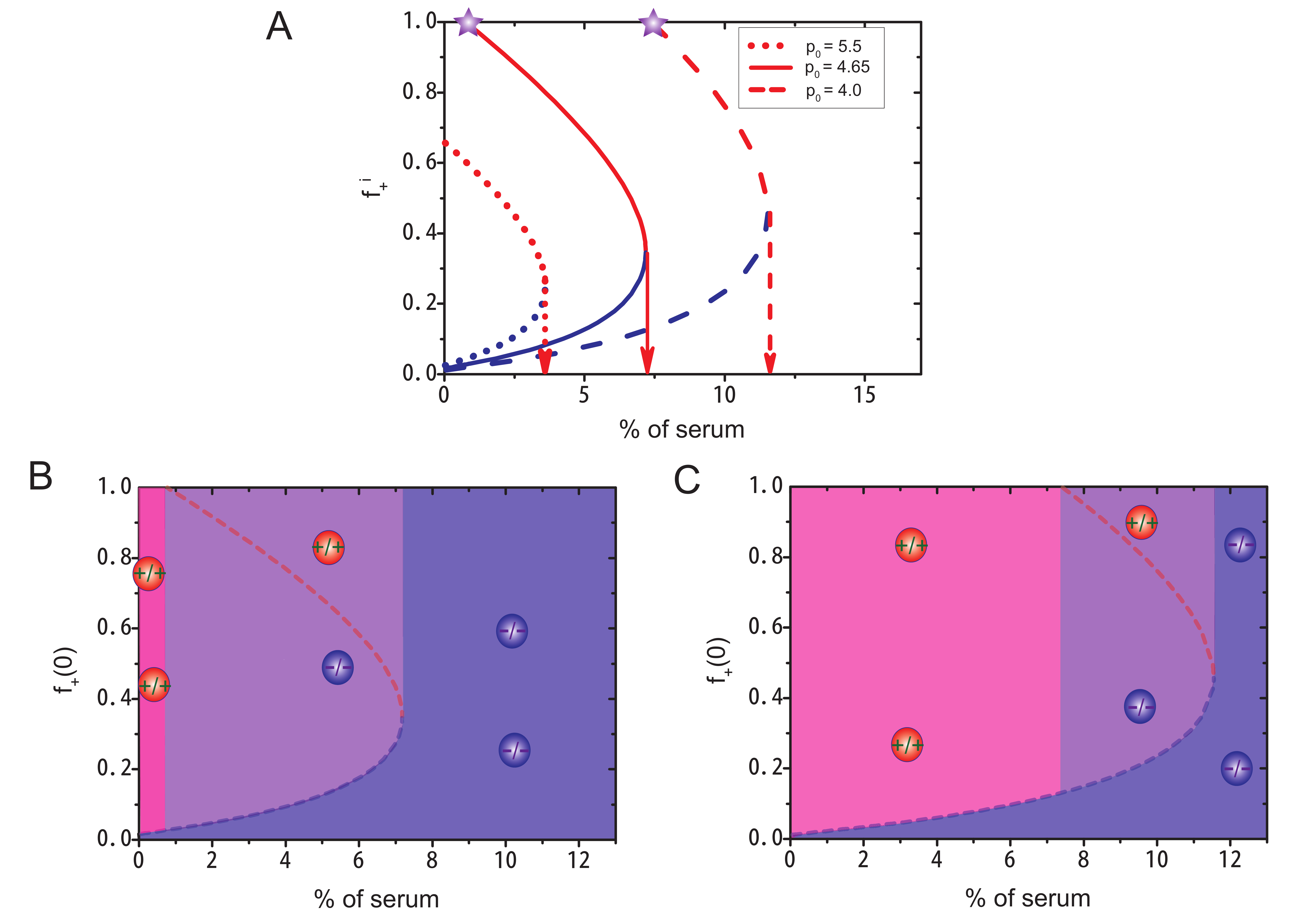}
\caption{\label{fig:price} (A) The internal equilibrium fractions ($f_{+}^{i}$, $i \equiv s, us$) of producers as a function of the levels of serum for different $p_{0}$ values. The fraction ($f_{+}^{s}$) at stable equilibrium are shown in red color while blue color indicates unstable equilibrium fractions ($f_{+}^{us}$). Arrows give the critical level of serum above which only non-producers exist. Purple stars represent the lowest level of serum at which producers and non-producers coexist in a stable equilibrium state. (B) and (C) Phase diagrams (initial fraction $f_{+}(0)$ vs \% of serum) with $p_{0} $= 4.65 and  $p_{0}$ = 4.00, respectively. Three stable phases are shown in these two figures. i) A homogeneous phase consisting of only producers (pink color). ii) A heterogeneous phase consisting of both producers and non-producers (purple color). The stable equilibrium fraction of producers is indicated by the dashed red line. iii) A homogeneous phase with of only non-producers (blue color).   The red and blue circles represent the producer and non-producer cells, respectively. Remarkably, for both $p_{0}$ values +/+ and -/- cells coexist in a narrow range of \% of serum.}
\end{figure}

Just as in Fig.~\ref{fig:phase},  we investigated the  internal equilibrium fraction $f_{+}^{i}$ of producers as a function of the level of exogenous resources but differing values of $p_{0}$.  The stable internal equilibrium fraction of producers is represented by the red curves while the blue curves report the unstable internal equilibrium fraction (Fig.~\ref{fig:price}A). A critical $p_{0}$-dependent concentration for exogenous resources is observed in Fig.~\ref{fig:phase} above which non-producers can maintain a stable homogeneous system irrespective of the fate  of the producers. The  value of the critical concentration (see the red arrows in Fig.~\ref{fig:price}A)  increases as $p_{0}$ decreases. The fitness of producers increases as they pay a much lower price (smaller $p_{0}$) for public goods production.  Therefore, additional exogenous resources have to be provided to enhance the competitive advantage of non-producers at small $p_{0}$.  We also observed another  critical value (indicated by the star symbols in Fig.~\ref{fig:price}A) for exogenous resources at relatively small $p_{0}$ values. Only stable homogeneous states (consisting of either producers or non-producers)  exist if the level of  exogenous resources fall below this critical value.  Interestingly, it becomes easier for the producer to establish a stable homogeneous system as $p_{0}$ decreases. On the other hand,  only a small increase of exogenous resources leads to a homogeneous tumor consisting of only non-producers, as $p_{0}$ takes on large values (see the dotted line in Fig.~\ref{fig:price}A). 

From Fig.~\ref{fig:price}A, we obtain the phase diagram in terms of the variables of the exogenous public goods concentration, and the initial fraction $f_{+}(0)$ of producers. Two examples,  shown in  Figs.~\ref{fig:price}B and \ref{fig:price}C  with $p_{0} = 4.65$, and $p_{0} = 4.0$, respectively, show the emergence of  three stable phases.  At low levels of exogenous resources and large $f_{+}(0)$, a homogeneous phase with only producers (shown in pink color) exists.  The second homogeneous phase, with only non-producers (shown in blue color), is easily accessible at high levels of exogenous resources. A heterogeneous phase representing the coexistence of both producer and non-producer cells (purple color) can be attained at intermediate levels of exogenous resources and large $f_{+}(0)$. By comparing Figs.~\ref{fig:price}B and \ref{fig:price}C, we find that  the parameter space for the non-producer to take over the whole system shrinks as $p_{0}$ decreases while it increases for the producer to dominate the system.  These figures also show that a heterogeneous system might be established  easily (within a larger parameter space, comparing the purple region in Figs.~\ref{fig:price}B and \ref{fig:price}C) if the producer pays a relatively high price for public goods production.  From these discussions,  we conclude that the price $p_{0}$ paid by producers greatly influences  the state of the tumor, especially the robustness of the heterogeneous state. It appears that one can design better treatment protocols for cancers composed of different  subpopulations by  regulation of certain parameters, such as $p_{0}$ discussed here.

\textbf{Cooperation among cancer subpopulations in {\it in vivo} experiments on glioblastoma:} The mechanism of cooperation and feedback described through the replicator equations might be operative in other cancers. In order to illustrate the applicability of our theory, we analyze the origin of ITH in Glioblastoma multiforme (GBM). It is known that GBM is the most common and aggressive primary brain cancer with poor prognosis. The five-year survival rate is less than five percent,  and most patients  live for  only a year following diagnosis and treatment~\cite{Gallego15}. The extensive presence of ITH in GBM is well-known at the genetic, molecular and cellular levels~\cite{Jung99,Bonavia11}. As in many other types of cancers, the mechanism for the origin of heterogeneity in GBM remains unclear, which is one reason in the poor design of  effective treatment. 

It is established~\cite{Hurtt92,Jaros92,Schlegel94} that chromosomal amplification of epidermal growth factor receptor gene (EGFR) is present in most cells of many primary GBMs. Another type of cell,  showing intragenic rearrangement of EGFR gene (with deletion of exons 2-7)  also frequently appear in the same tumor~\cite{Nishikawa94}.  The coexistence of these two types of cells with differing expressions of the growth factor receptor leads to a worse prognosis of GBM~\cite{Shinojima03,Heimberger05} than would be the case when the cell (with EGFR gene rearrangement) is absent. 

Recently, an experiment~\cite{Inda2010} studied  the interactions between tumor cells with amplified levels of EGFR (referred to  wt cells) and cells with rearrangement of EGFR gene (called  $\Delta$ cells) within a neoplasm.  
It is found that the total size of the tumors (after 12 days of orthotopic injection) is much larger if a mixture of  wt and $\Delta$ cells are injected into one mouse simultaneously than when they are injected into two mice separately. This finding shows that these two types of cells cooperate with each other to promote growth of the tumor.  The producer ($\Delta$) cells secretes certain factors like Interleukin-6(IL-6) and/or Leukemia inhibitory factor (LIF), which enhance the proliferation and inhibit apoptosis of tumor cells~\cite{Inda2010,Heinrich03}.  The system composed of wt and $\Delta$ cells is analogous to the one considered in the previous sections with IL-6 and/or LIF playing the role of the public goods. Therefore, we can apply our theory to investigate the consequence of cooperation between these two cell types in GBM. 

\begin{figure}
\vspace*{-0.1cm}
\hspace*{-0.2cm}
\centering
\includegraphics[clip,width=1.0\textwidth]{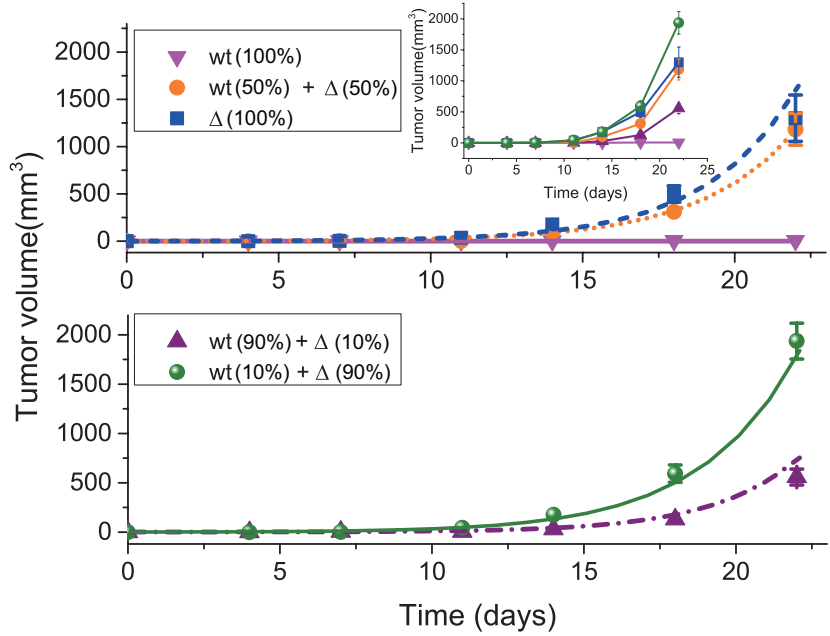}
\caption{\label{fig:tumorvolume} The evolution of tumor size as a function of time (in days) in glioblastoma with only wt cells, mutated $\Delta$ cells, or mixture of these two types of cells (wt +$\Delta$) injected into nude mice. The symbols represent experimental data. The parameter values ($a = 68.4, b=0.946$ and $p_{0}=0.651$) in the model are obtained by fitting the theory to experimental data (details in the SI) with 100$\%$ wt, 100$\%$ $\Delta$ and 50$\%$ for each type of cells (upper panel). Lower panel: The purple and green curves are theoretical predictions with 10$\%$, and 90$\%$ of $\Delta$ cells,  which both agree quantitatively with experimental observations. Error bars represent the standard error of the mean in experiments. The complete experimental data are shown in the inset (the labels are the same as in the main figure).}
\end{figure}

In the experiment~\cite{Inda2010}, the evolution of the tumor size was measured over a wide range of conditions. A fixed total number of tumor cells (with differing fractions of wt and $\Delta$ cells) are injected into nude mice,  and then the increase in the tumor volume is measured after different periods of time (see the inset in Fig.~\ref{fig:tumorvolume}).  Without $\Delta$ cells, it is difficult for the wt cells to induce tumor in nude mice,  as illustrated by the pink upside down-triangles. However, the $\Delta$ cell alone gives rise to tumors at a rapid growth rate, as illustrated by the blue squares in Fig.~\ref{fig:tumorvolume}. As long as a small fraction of $\Delta$ cells is  injected into the mice together with wt cells, fast growing tumors are induced in mice (see the purple up-triangles in Fig.~\ref{fig:tumorvolume}).  The tumor grows faster as the fraction of $\Delta$ cells in the total injected cells increases from 0, 10, 50, to 90$\%$ as shown in the inset of Fig.~\ref{fig:tumorvolume}. It is also remarkable that the tumors grow even faster as the injected cells are composed of 10$\%$ wt and 90$\%$ $\Delta$ cells (green spheres) than is the case when 100$\%$ of cells are $\Delta$ cells (blue squares), which again shows that  cooperation between the two cell types leads to enhanced growth rate.

\begin{figure}
\vspace*{-0.1cm}
\hspace*{-0.2cm}
\centering
\includegraphics[clip,width=1.0\textwidth]{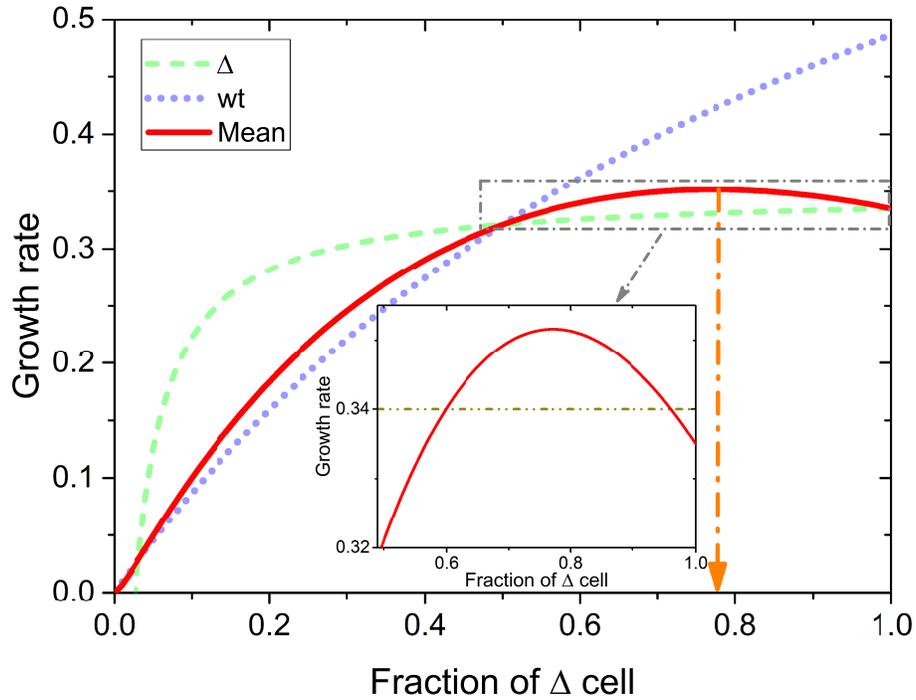}
\caption{\label{fig:Meanvelocity} Predictions of the mean growth rate of the GBM tumor cells as a function of the fraction of the $\Delta$ cell. The green dashed line gives the growth rate of $\Delta$ cells, and the purple dotted line shows the rate of wt cells. The average growth rate of the whole population with both types of cells are given by solid red line.  A maximum is observed at a value $f_{+}\approx 0.77$  (the orange arrow). The inset shows a zoom-in of the dash-dotted line rectangle. The values for the parameters $a, b$, and $p_{0}$  in this figure are the same as the ones used in Fig.~6.   The unit for growth rate is day$^{-1}$.}
\end{figure}

The experimental observations~\cite{Inda2010} can be quantitatively explained by the theoretical model developed here. By using the three growth curves (0$\%$, 50$\%$, and 100$\%$ $\Delta$ cells) for the tumors,  illustrated in the upper panel of Fig.~\ref{fig:tumorvolume}, we determined all the free parameters needed in the model (see SI for details). Then, the evolution of the tumor size at differing conditions can be predicted. The theoretical predictions for the tumor growth at 10$\%$, and 90$\%$ of $\Delta$ cells agree quantitatively with experimental observations, as shown in the lower panel of Fig.~\ref{fig:tumorvolume}. To further explain the growth curves shown in the inset of Fig.~\ref{fig:tumorvolume}, we plotted the growth rate of tumors as a function of the fraction of $\Delta$ cells (see the solid red line in Fig.~\ref{fig:Meanvelocity}). The growth rates for wt, $\Delta$ cells in the tumor are illustrated by dotted and dashed lines in Fig.~\ref{fig:Meanvelocity}, respectively. From this figure, it follows that the tumor growth rate increases as the fraction of $\Delta$ cells increases, reaching a maximum value  in the middle (0.77,  marked by the orange arrow). Subsequently, the rate starts to decrease. This behavior is similar to the experimental data in the inset of Fig.~\ref{fig:tumorvolume} and is also found for pancreatic cancer cells, as discussed here and discovered by AFC~\cite{Archetti15}.  

We also found that the glioblastoma with ITH is quite stable irrespective of the initial composition (see  Fig.~S5A). Our results explain the finding that frequently  the wt cells and $\Delta$ cells coexist in GBM, and provides an explanation for the poor prognosis due to the quick recovery of the fast growing state as long as a small fraction $f_{+}$ of $\Delta$ cells is present. Therefore, the stability of such a heterogeneous tumor needs to be eliminated in order to improve the survival rates of GBM patients.  From the discussion above, it follows that the supply of exogenous pubic goods can influence  cooperation between two different populations sharing one public good.  By adding exogenous cytokines to the model (see Eqs.~(S.7) and (S.8) in SI), a stable homogeneous system composed of only  wt cells could be reached (see Fig.~S5B) irrespective of the initial fraction of the producer, $\Delta$ cells. Such a tumor would stop growing after removal of exogenous cytokines as  wt cells alone cannot sustain tumor growth, as observed in experiments~\cite{Inda2010}. If practice, this might provide a treatment strategy for GBM, and perhaps other types of cancers dominated by ITH  due to the interactions among different cancer cells.

\section*{Discussion}

In this article, we investigated the interactions between two distinct subpopulations frequently observed in many cancers, which is a manifestation of heterogeneity.  We uncovered a general mechanism for the establishment of  a stable heterogeneous system consisting of producers and non-producer cells as a function of a number of experimentally controllable parameters. 
The tragedy of the commons would be expected as the public goods are shared equally among both the populations.  However, a stable heterogeneous state arises if the producer can obtain the public goods more efficiently than the non-producer. 
 Most importantly, the emergence of these scenarios require the fitness of the two players must be a non-linear function of the public goods. Otherwise, only an unstable heterogeneous system  can be established. In addition, the cost to benefit ratio is a critical factor in determining the establishment of a stable coexisting state. In the experiments~\cite{Archetti15}, the benefit is adjusted by changing the amount of serum while the cost of public good production is a constant. However, Archetti et al changed the cost instead of benefit to study the cooperation and competition of the two types of cells in their model. This is due to the complex payoff function assumed in the AFC model. In our model, the benefit of public goods is separated into two parts (endogenous and exogenous) naturally, while the cost is a constant. Therefore, we can investigate the influence of benefits on the cell cooperation and competition directly, as realized in experiments. In addition, our formulation is sufficiently general that we could test the effects of all other experimentally accessible parameters in order to assess the ranges of parameters that produce coexistence between producer and non-producer cells, as illustrated for the specific case in Fig. 5B.
 
 We also found that it is relatively easy to establish cooperation  and form a stable diversified or heterogeneous state in harsh conditions than in resource-abundant conditions. Such a phenomenon is quite common in biological systems~\cite{Korolev2014,Hoek2016}.  The price paid by the producers also strongly influences the cooperation between the two players. Higher price can decrease the demand for exogenous resources in order to establish cooperation and might also expand cooperation to a wide parameter range.

Frequently in many cancers a minor subclone could support the growth of the whole tumor consisting of many different subpopulations~\cite{Mullighan2008,Johnson2014,Morrissy2016}. It is easy to detect the genotype of dominant subclones, which would be considered as the target  in later treatments. However, if a minor subclone escapes detection then it  could survive,  promoting a faster and more aggressive tumor growth caused by the competitive release~\cite{greaves2012clonal}. Therefore, it is essential to learn the composition of a heterogeneous tumor, and also the interactions among different subclones before efficacious treatment can be formulated for the patients.  

For cancers with producer and non-producer cells discussed here, it might be prudent to feed these cells  instead of depriving them of nutrients so that  competition between different subclones is promoted. Then, an effective treatment can be implemented  as the system transits from a stable heterogeneous population to a homogeneous population.  We have illustrated this concept using an experiment involving glioblastoma.  This idea is reminiscent of  another concept in cancer therapy, the tumor vasculature normalization~\cite{jain2005normalization}. The tumor vasculature is quite abnormal, which leads to heterogeneous tumor blood flow. Therefore, many tumor cells cannot get access to blood vessels and live under pressure such as hypoxia and acidosis, thus inducing genome instability and high intratumor heterogeneity~\cite{bristow2008hypoxia}. Temporal normalization of tumor vasculature can reduce the microenvironment pressure on tumor cells and also increase the drug delivery efficiency. Hence, it can increase the conventional therapy efficacy if both procedures are scheduled carefully~\cite{jain2005normalization}. Similarly, the new idea proposed here could be combined with traditional therapies, such as surgery and chemotherapy,  to reduce the risk and drug resistance but increase the therapy efficacy.   In addition, similar public goods dilemma has been observed in many systems, such as microbial colonies, insect communities, and human society~\cite{dobata2013public,drescher2014solutions,kaul1999global}. The mechanism proposed here for the establishment of a heterogeneous population is quite general, and could in principle be applied to these systems.  It will be most interesting to extend our model to the case beyond two species, which is more prevalent in nature~\cite{levine2017beyond}. It would be fruitful to consider different mechanisms of ITH in order to account for  complex origins of ITH~\cite{Almendro13}.

\vspace{6mm}
\noindent {\bf SUPPLEMENTAL INFORMATION}\\
\noindent Supplemental Information including five figures and detailed methods can be found with this article online.\\

\noindent {\bf ACKNOWLEDGMENTS} \\
\noindent  We are grateful to Abdul N Malmi-Kakkada and Sumit Sinha for discussions and comments on the manuscript.\\

\noindent {\bf AUTHOR CONTRIBUTIONS}\\
\noindent X. L. and D. T. conceived and designed the project, and co-wrote the paper. X. L. performed the research.\\

\noindent {\bf Competing interests}\\
\noindent We declare we have no competing interests.\\

\noindent {\bf Funding}\\
\noindent This work is supported by the National Science Foundation (PHY 17-08128 and CHE 16-32756), and the Collie-Welch Chair through the Welch Foundation (F-0019).

\bibliographystyle{unsrt}

\bibliography{cancerheterogeneitytexrevise1}

\newpage

\newpage


\newpage

\newpage

\newpage

\newpage

\newpage

\end{document}